\documentclass[aps,prd,amsfonts,amssymb,amsmath,showpacs]{revtex4}

\usepackage{graphicx}
\usepackage{dcolumn}
\usepackage{bm}

\begin{document}

\nopagebreak

\title{Thermodynamics of de Sitter black holes
with a conformally coupled scalar field}

\author{Anne-Marie Barlow}
\author{Daniel Doherty}
\author{Elizabeth Winstanley}
    \email{E.Winstanley@sheffield.ac.uk}
\affiliation{Department of Applied Mathematics, The University of Sheffield,
Hicks Building, Hounsfield Road, Sheffield, S3 7RH, United Kingdom.}

\date{\today}

\begin{abstract}
We study the thermodynamics of de Sitter black holes with a conformally coupled
scalar field.
The geometry is that of the ``lukewarm'' Reissner-Nordstr\"om-de Sitter
black holes, with
the event and cosmological horizons at the same temperature.
This means that the region between the event and cosmological horizons
can form a regular Euclidean instanton.
The entropy is modified by the non-minimal coupling of the scalar field to the geometry,
but can still be derived from the Euclidean action, provided suitable
modifications are made to deal with the electrically charged case.
We use the first law as derived from the isolated horizons formalism to compute
the local horizon energies for the event and cosmological horizons.
\end{abstract}


\maketitle

\section{Introduction}
\label{sec:intro}

Within the subject of black hole thermodynamics, the study of black holes with
scalar or dilaton fields is an important area, in which new features arise.
For example, there is a flat-space black hole solution (the BBMB black hole \cite{BBMB})
with a conformally coupled scalar field, which has zero temperature but infinite
entropy \cite{zaslavskii}.
In \cite{zaslavskii}, a modified action principle is formulated which is able to
provide well-defined thermodynamics for a number of black holes with an infinite
entropy, although, interestingly, this approach fails for the BBMB black hole.

In this paper we study the thermodynamics of the corresponding de Sitter black holes
with a conformally coupled scalar field (which we shall refer to as the
MTZ black holes) \cite{mtz}.
Here the geometry is that of the ``lukewarm'' Reissner-Nordstr\"om-de Sitter (RNdS) black holes
\cite{romans}, but, unlike the BBMB black holes, the scalar field is regular on
and outside the event horizon.
The thermodynamics of RNdS black holes without a scalar field has been studied by
several authors \cite{romans,mann,cai}.
In the lukewarm case, the event and cosmological horizons are at the same
temperature, which enables a regular Euclidean instanton to be formed from the region
between the event and cosmological horizons \cite{mellor}.
The Euclidean action for the lukewarm RNdS black hole and its relationship to the entropy
has also been studied \cite{mann,cai}.
Our interest in this paper is therefore to determine the effect of the non-minimal coupling
of the scalar field matter to the black hole.
The thermodynamics of a related de Sitter black hole with a {\em {minimally}} coupled
scalar field is discussed in \cite{zlosh}.

The outline of this paper is as follows.
In section \ref{sec:mtz} we briefly review the salient features of the MTZ black holes.
We compute the temperature and entropy of these black holes in section \ref{sec:thermo}
and discuss elementary features of these
(see also \cite{milan} for further discussion).
In particular, we will see the significant effect of the conformally coupled scalar field
on the entropy of the system.
We examine the entropy in more detail in section \ref{sec:action}, by computing the
Euclidean action for the regular instanton \cite{mellor}.
This is straightforward for magnetically charged black holes, but there is a subtlety
for the electrically charged case, which also arises for RNdS black holes without a scalar
field \cite{mann}.
Since we are focusing on the region between the horizons, which does not include infinity,
we do not compute the global gravitational mass, but instead, in section \ref{sec:firstlaw}
study the local horizon energy defined through the first law of isolated horizons
\cite{ashtekar}.
Our conclusions are presented in section \ref{sec:conc}.

\section{The MTZ black holes}
\label{sec:mtz}

The MTZ black holes \cite{mtz} are solutions of the field equations arising from
the action
\begin{equation}
I_{L}  =   \int _{{\cal {M}}} d^{4}x \sqrt{-g} \left[
\frac {1}{16\pi } \left( R-2\Lambda \right)
- \frac {1}{2}
g^{\mu\nu} \partial_{\mu} \phi \partial_{\nu} \phi
-\frac{1}{12}R\phi^{2}
 - \alpha\phi^{4}
-\frac{1}{16\pi}F^{\mu\nu}F_{\mu\nu} \right] ;
\label{eq:action}
\end{equation}
where $\alpha $ is the coupling constant.
Here and throughout this paper, we use units in which $c=G=k_{B}=\hbar =1$.
We will consider both charged and electrically neutral
black holes, so for neutral black holes,
the electromagnetic field is absent.
In both cases, the black hole geometry is described by the
RNdS metric:
\begin{equation}
\label{eq:metric}
ds^{2}=-N(r)dt^{2}+ N(r)^{-1}dr^{2}+ r^{2}\left(
d\theta^{2}+\sin^{2}\theta \, d\varphi^{2}\right) ;
\end{equation}
where
\begin{equation}
N(r)= \left(1-\frac{M}{r}\right)^{2}
-\frac {\Lambda }{3} r^{2} .
\end{equation}
The black holes have inner, event and cosmological horizons at
values of the radial co-ordinate $r$ given by, respectively,
\begin{eqnarray}
r_{-} & = & \frac {l}{2} \left[ -1 + {\sqrt {1+ \frac {4M}{l}}}
\right] ;
\nonumber \\
r_{+} & = & \frac {l}{2} \left[ 1 - {\sqrt {1 - \frac {4M}{l}}}
\right] ;
\nonumber \\
r_{++} & = & \frac {l}{2} \left[ 1 + {\sqrt {1 - \frac {4M}{l}}}
\right] ;
\label{eq:hor}
\end{eqnarray}
where $l={\sqrt {3/\Lambda }}$.
From (\ref{eq:hor}), it is clear
that the solution is defined only for $0<M<M_{{\rm {max}}}=l/4$.
When $M=0$, the metric (\ref{eq:metric}) reduces to de Sitter space and there
is only a cosmological horizon.
In the other limit, $M=M_{{\rm {max}}}=l/4$, the event and cosmological horizons
coincide, leaving a charged Nariai solution.
The Penrose diagram for the generic case can be found in \cite{mellor}.

The form of the scalar field
depends on whether we are studying neutral or charged solutions.
In the case with no electromagnetic
field, there is a solution only if the coupling constant $\alpha $
is given by
\begin{equation}
\alpha = -\frac {2\pi \Lambda }{9} ;
\end{equation}
and then the scalar field takes the form
\begin{equation}
\phi (r) = \frac {{\sqrt {3}}M}{{\sqrt {4\pi }}(r-M)} ;
\label{eq:phineut}
\end{equation}
which has a pole at $r=M<r_{+}$, lying inside the event horizon.

For electrically charged black holes, the only non-vanishing component of the
electromagnetic field is
\begin{equation}
F_{tr}=-\partial_{r}A_{t}=\frac{Q}{r^{2}};
\label{eq:electric}
\end{equation}
where the charge-to-mass ratio is given by a quantity $K\in [0,1]$, defined as
\begin{equation}
\label{eq:qm}
K ^{2} =  \left( \frac{Q}{M} \right)^{2}
= \left( 1+\frac {2\pi \Lambda }{9\alpha } \right) ;
\end{equation}
and the scalar field in this case is
\begin{equation}
\label{eq:phicharged}
\phi(r)= {\sqrt{ -\frac{\Lambda}{6\alpha}}} \frac {M}{r-M}
= \frac {{\sqrt {3(M^{2}- Q^{2})}}}{\sqrt {4\pi }(r-M)}.
\end{equation}
This latter solution only exists provided $\alpha$ satisfies the
bound
\begin{equation}
\alpha < - \frac {2\pi \Lambda }{9}.
\end{equation}
If we set $K=0$, then we recover the neutral black hole scalar field (\ref{eq:phineut}),
while the value $K=1$ corresponds to the usual RNdS black hole with $Q=M$.
In this latter case, we have effectively taken the limit $\alpha \rightarrow -\infty $,
so that the scalar field (\ref{eq:phicharged}) is zero.
In addition to the electrically charged solutions described in \cite{mtz}, there
are also magnetically charged solutions, with the scalar field still given by
(\ref{eq:phicharged}), but with the electromagnetic field having a single non-zero component:
\begin{equation}
F_{\theta \phi } = Q \sin \theta .
\label{eq:magnetic}
\end{equation}
The neutral and electrically charged MTZ solutions were shown to be unstable in \cite{harper},
and we conjecture that the magnetically charged black holes are also classically unstable.
The fact that these black holes are classically unstable does not render their
thermodynamics uninteresting.
For example, the BBMB black hole is also classically unstable, yet the study of its
thermodynamics has revealed deeper insights into the subject \cite{zaslavskii}.

\section{Temperature and Entropy}
\label{sec:thermo}

Although the geometry (\ref{eq:metric}) of the black holes we
are studying in this note is that of a standard
RNdS black hole, the presence of the conformally coupled scalar field affects the
thermodynamical quantities significantly.

The temperature of the black holes is unaffected by the conformally coupled scalar
field, however, and given by the usual Hawking formula:
\begin{equation}
T _{\Delta }
= \frac {\kappa _{\Delta }}{2\pi }= \frac {1}{4\pi } \left| N' (r_{\Delta }) \right| .
\label{eq:temp}
\end{equation}
As observed in \cite{mellor,romans,cai},
the temperature given by (\ref{eq:temp}) is the same for the event and cosmological horizons:
\begin{equation}
T= \frac {1}{2\pi l} {\sqrt {1-\frac {4M}{l}}} .
\label{eq:MTZtemp}
\end{equation}
Therefore, we are considering ``lukewarm'' black holes in the terminology of \cite{romans}.
Note that $0\le 2\pi T \le 1/l$.
When $M=0$, the temperature $T$ is simply the de Sitter temperature $T=1/2\pi l$, while if
$M=M_{{\rm {max}}}=l/4$, we have $T=0$.
Unlike the BBMB black hole, the temperature is non-zero in general.

On the other hand, the entropy of the black hole is no longer given by the usual area
law, but acquires a multiplicative factor due to the conformally coupled scalar
field \cite{iyer,ashtekar} (see also \cite{zanelli} for the 2+1-dimensional case):
\begin{equation}
S_{\Delta } = \pi r_{\Delta }^{2} \left( 1- \frac {1}{6} \phi (r_{\Delta })^{2} \right) .
\label{eq:entropy}
\end{equation}
There are a number of ways of deriving this result.
Firstly, there is the approach due to Iyer and Wald \cite{iyer}, where the entropy of a black
hole in a theory with a very general action is derived using a Noether current approach to
the first law.
This method gives the entropy to be:
\begin{equation}
S= -2 \pi  \int _{\Sigma } E^{\mu \nu \alpha \beta } \epsilon _{\mu \nu }
\epsilon _{\alpha \beta } ;
\end{equation}
where the integration is performed over the bifurcation two-sphere of the event horizon
$\Sigma $, with binormal $\epsilon _{\mu \nu }$, and $E^{\mu \nu \alpha \beta }$ is
the functional derivative of the Lagrangian $L$ of the theory with respect to
$R_{\mu \nu \alpha \beta }$, treated as a variable independent of the metric $g_{\mu \nu }$.
Applying this formula to our action (\ref{eq:action}) gives the entropy (\ref{eq:entropy}).
The entropy (\ref{eq:entropy}) has also been rederived recently using the isolated horizons approach
to deriving the first law \cite{ashtekar}.
We shall employ this approach in section \ref{sec:firstlaw} to compute the local energies
of the event and cosmological horizons.
We will also discuss, in section \ref{sec:action}, a third alternative, following a method of
Kallosh et al \cite{kallosh}.
In the case of the BBMB black hole, the formula (\ref{eq:entropy}) gives an infinite answer
because the scalar field is not regular on the event horizon.
This is discussed in detail in \cite{zaslavskii}.

Although the scalar field $\phi $ is constant over each of the horizons,
the multiplicative factor in (\ref{eq:entropy}) has a significant qualitative effect on the entropy.
For the black hole event horizon at $r=r_{+}$, we find the entropy to be:
\begin{equation}
S_{+}= \pi l^{2}
\left[
-{\sqrt {1-\frac {4M}{l} }} + \frac {K^{2}}{2} \left(
1 - \frac {2M}{l} +
{\sqrt {1- \frac {4M}{l}}}
\right)
\right] ,
\label{eq:Sr+}
\end{equation}
while for the cosmological horizon at $r=r_{++}$,
\begin{equation}
S_{++} =
\pi l^{2}
\left[
{\sqrt {1-\frac {4M}{l} }} + \frac {K^{2}}{2} \left(
1 - \frac {2M}{l} -
{\sqrt {1- \frac {4M}{l}}}
\right)
\right] .
\label{eq:Sr++}
\end{equation}
The entropies (\ref{eq:Sr+},\ref{eq:Sr++}) can also be written compactly in
terms of the temperature $T$:
\begin{eqnarray}
S_{+} & = &
\frac {\pi }{4}\left[
-8\pi l^{3} T + K^{2}l^{2} \left( 1 + 2 \pi l T \right) ^{2}
\right] ;
\nonumber \\
S_{++} & = &
\frac {\pi }{4}\left[
8\pi l^{3} T + K^{2}l^{2}  \left( 1 - 2 \pi l T \right) ^{2}
\right] .
\label{eq:SvT}
\end{eqnarray}
The total entropy of the space-time is given by the sum of the entropies arising from
the event and cosmological horizons:
\begin{equation}
S_{\rm {total}} = S_{+} + S_{++}
= \frac {\pi }{2} K^{2}l^{2} \left( 1 + 4 \pi ^{2} l^{2} T^{2} \right)
= \pi  K^{2} l^{2} \left( 1 - \frac {2M}{l} \right).
\label{eq:Stotal}
\end{equation}

For the neutral black holes ($K=0$), we see that the black hole event horizon
entropy $S_{+}$ is always negative, and further, that $S_{+}=-S_{++}$, giving
a zero total entropy.
This is despite the fact that the temperature is non-zero.
The other limit of interest is when $K=1$, in which case the entropies
$S_{\rm {total}}$, $S_{+}$ and $S_{++}$
all reduce to the known values for the ordinary RNdS black holes.
The entropy of the cosmological horizon, $S_{++}$, is always positive, but the
entropy of the black hole event horizon, $S_{+}$ is negative unless $T$ is sufficiently
small and $K^{2}>0$.

The fact that the entropy of the black hole event horizon is
negative for most of the available parameter space is very interesting.
The negative entropy arises because of the non-minimal coupling between the
geometric curvature and the scalar field.
Similar negative entropies can occur for black holes in higher-derivative theories
(see, for example, \cite{cvetic}).
This may indicate some fundamental thermodynamical instability, in parallel with
the known classical instability of these solutions \cite{harper}.
However, we should stress that the total entropy of the system is always positive or
zero.

\begin{figure}
\begin{center}
\includegraphics[angle=270,width=10cm]{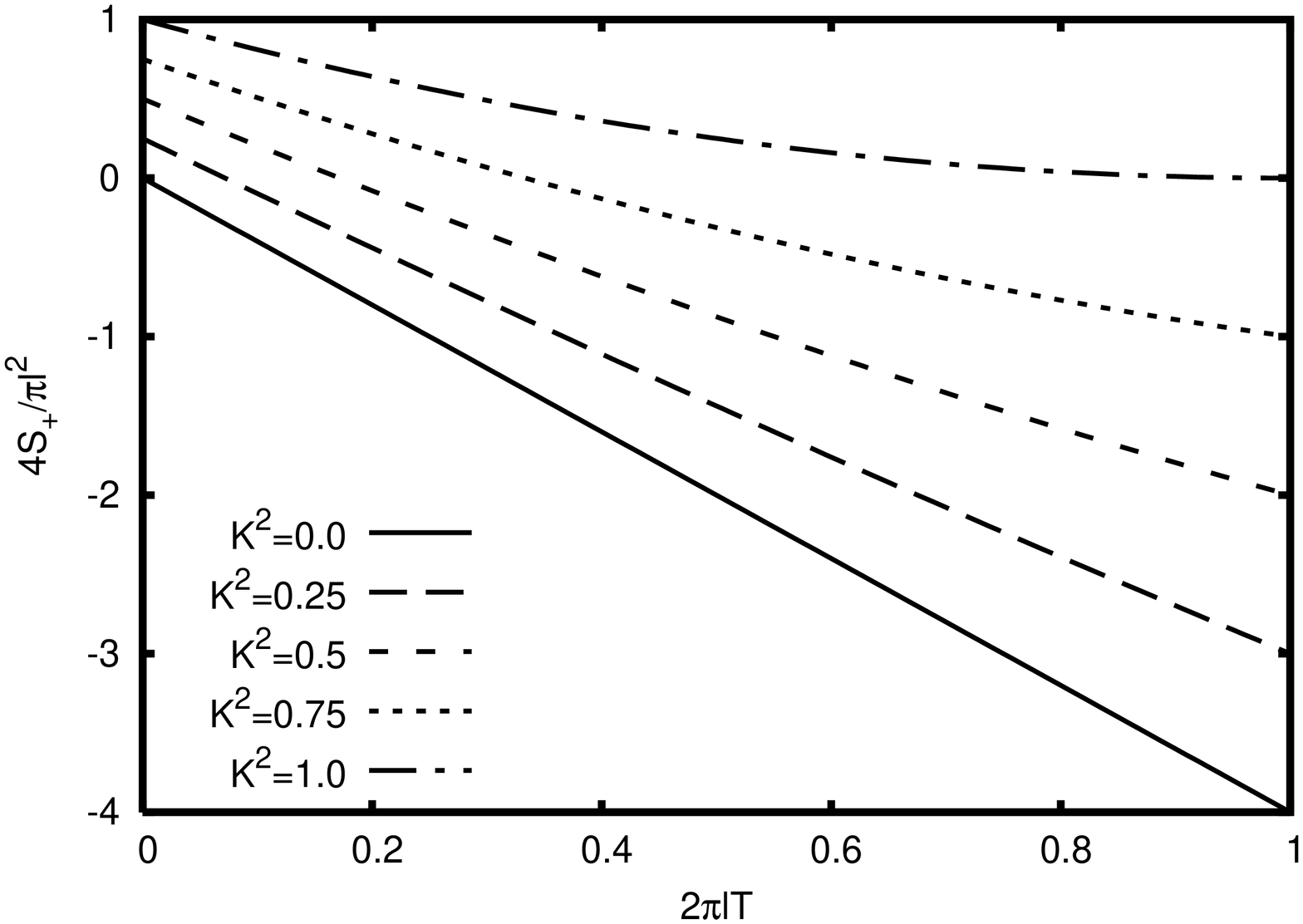}
\includegraphics[angle=270,width=10cm]{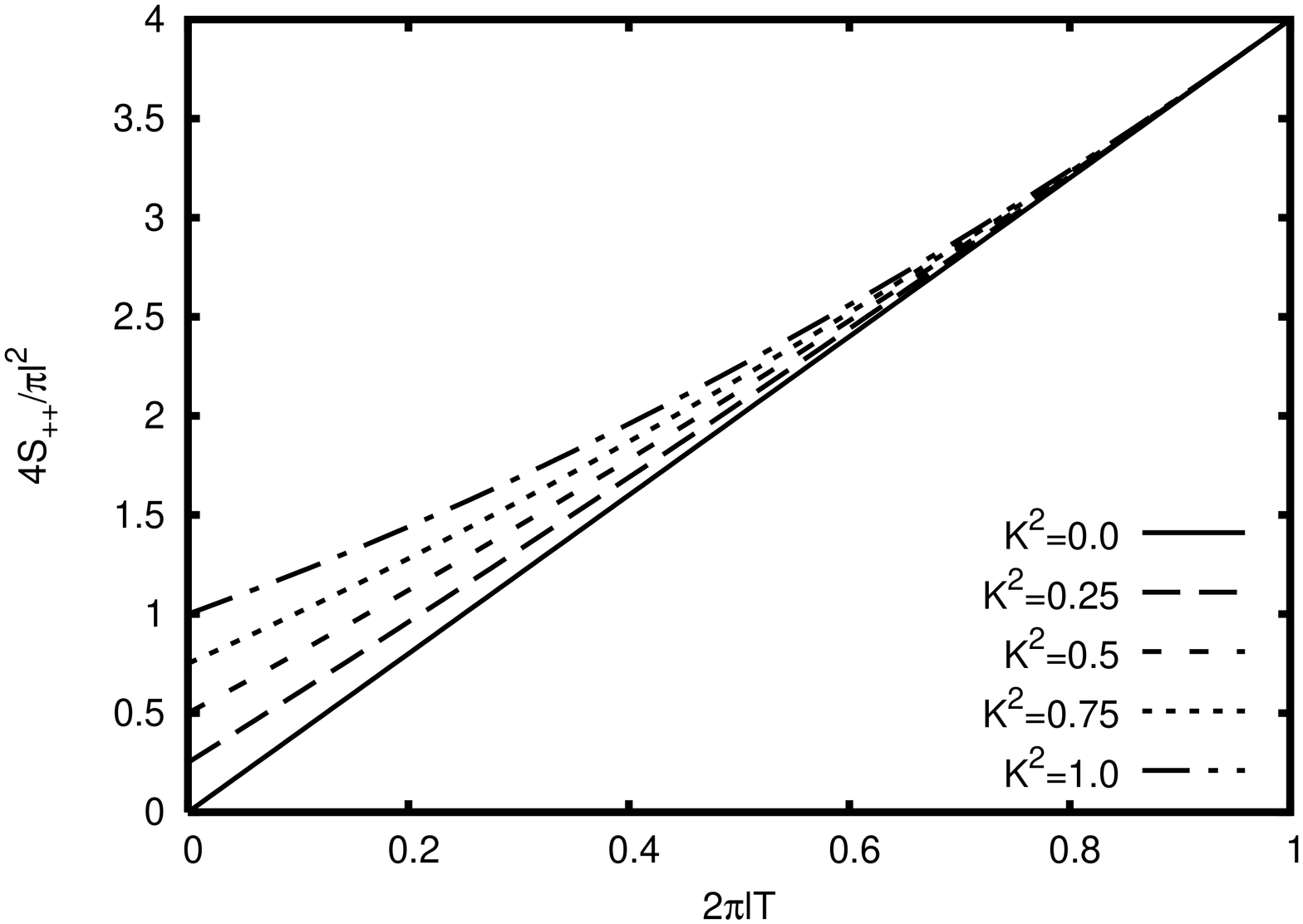}
\end{center}
\caption{Graph of the entropies of the event horizon, $S_{+}$ (top) and
cosmological horizon, $S_{++}$ (bottom) as functions of the temperature $T$,
for various values of the charge/mass ratio.}
\label{fig:ST}
\end{figure}

We are unable to regard $Q$ as a free parameter, as it is fixed by $M$ and the coupling
constants of the theory (\ref{eq:qm}).
Therefore we plot in Figure \ref{fig:ST} the entropies $S_{+}$ and $S_{++}$ as functions
of the temperature $T$ for fixed values of the charge/mass ratio $K$.
It is clear from Figure \ref{fig:ST}, or by direct differentiation of (\ref{eq:SvT}),
that the derivatives of $S_{+}$ and $S_{++}$ are finite and non-zero,
for all values of $K$ and $T$, except that $\partial S_{+}/\partial T=0$ when $K=1=T$.
This means that there are no phase transitions, in contrast to the situation when
there is no scalar field \cite{davies}.
The entropies $S_{+}$ and $S_{++}$ are discussed in more detail in \cite{milan}.

\section{Entropy and Euclidean Action}
\label{sec:action}

As the event and cosmological horizons are at the same temperature (\ref{eq:MTZtemp}),
a regular Euclidean instanton can be formed from the region
between the horizons \cite{mellor}.
We complexify the time co-ordinate $t$ and introduce a new real co-ordinate $\tau $
by $\tau = -it$.
The resulting Euclidean section has metric
\begin{equation}
ds^{2}=N(r)d\tau ^{2}+ N(r)^{-1}dr^{2}+ r^{2}\left(
d\theta^{2}+\sin^{2}\theta \, d\varphi^{2}\right) ,
\label{eq:Emetric}
\end{equation}
and by taking the Euclidean time $\tau $ to have
period $1/T$,
a regular instanton ${\cal {M}}$ is formed with topology
$S^{2} \times S^{2}$ \cite{mellor}.

We now consider the Euclidean action
\begin{eqnarray}
I_{E} & = &
 - \int _{{\cal {M}}} d^{4}x \sqrt{g} \left[
\frac {1}{16\pi } \left( R-2\Lambda \right)
- \frac {1}{2}
g^{\mu\nu} \partial_{\mu} \phi \partial_{\nu} \phi
-\frac{1}{12}R\phi^{2}
 - \alpha\phi^{4}
-\frac{1}{16\pi}F^{\mu\nu}F_{\mu\nu} \right]
\nonumber \\ & &
-\int _{\partial {\cal {M}}} d^{3} x {\sqrt {h}}
\left[ 1 - \frac {1}{6} \phi ^{2} \right] K ;
\label{eq:IE}
\end{eqnarray}
where $K$ is the trace of the second fundamental form of the boundary $\partial {\cal {M}}$.
The boundary term is a generalization of the Gibbons-Hawking \cite{gh} term for
a conformally coupled scalar field, and is required to ensure that the variational
principle is well-defined.
Since we are considering the instanton formed from the region between the event and
cosmological horizons, we do not need to include additional counterterms to
deal with the boundary at infinity.
The instanton we are considering is of finite spatial extent, and so
we are using the microcanonical ensemble \cite{york}.
From this it follows that the Euclidean action (\ref{eq:IE}) should be
equal to minus the total entropy of the system.

The Euclidean instanton has no boundary, so the boundary term in (\ref{eq:IE}) does
not contribute.
We begin by looking at the magnetically charged black holes, since
their Euclideanization is straightforward, with the magnetic field
still given by (\ref{eq:magnetic}).
A simple calculation reveals that
\begin{equation}
I_{E} = - \pi K^{2} l^{2} \left( 1 - \frac {2M}{l} \right)
\label{eq:IEmag}
\end{equation}
in this case.
We can immediately see that the Euclidean action is equal to minus the total entropy
(\ref{eq:Stotal}) for the neutral and magnetically charged black holes.
This is in agreement with the results for the lukewarm RNdS black holes without a scalar
field \cite{mann}.

The situation for the electrically charged black holes is more subtle.
We could follow \cite{mellor} and complexify the electric charge so that
the electric field (\ref{eq:electric}) becomes
\begin{equation}
F_{\tau r}=\frac{{\hat {Q}}}{r^{2}},
\label{eq:FEel1}
\end{equation}
where ${\hat {Q}}=iQ$ is {\em {real}}.
This gives the Euclidean action to be exactly the same as in the magnetically charged
case (\ref{eq:IEmag}).
However, such a transformation does not preserve electromagnetic duality
\cite{hawkross}, and, furthermore, in this case the
metric (\ref{eq:Emetric}) no longer satisfies the Einstein equations.
If we wish to have (\ref{eq:Emetric}) as a solution of the Einstein equations
and also retain electromagnetic duality, then we keep the parameter $Q$ {\em {real}},
leading to an {\em {imaginary}} electric charge in the Euclidean section, with
electric field \cite{hawkross,mann,cai}:
\begin{equation}
F_{\tau r}=\frac{iQ}{r^{2}}.
\label{eq:FEel2}
\end{equation}
In this case the Euclidean action (\ref{eq:IE}) turns out to be
\begin{equation}
I_{E} = - \pi K^{2} l^{2} \left( 1 - \frac {4M}{l} \right) .
\label{eq:IEel}
\end{equation}
The difference between the two answers (\ref{eq:IEmag}) and (\ref{eq:IEel})
is due to the fact that
\begin{equation}
F^{\mu\nu}F_{\mu\nu} = -\frac {2Q^{2}}{r^{4}}
\end{equation}
when we have the electric field (\ref{eq:FEel2}), but has the opposite sign for
the electric field (\ref{eq:FEel1}) (and for the magnetically charged black holes).

The situation we have is identical to that for the lukewarm RNdS black holes
without a scalar field \cite{mann,cai}, namely
the action (\ref{eq:IEel}) does not equal minus the total entropy.
However, as discussed in \cite{mann,york}, we are dealing with an instanton manifold
of finite spatial extent
and so our action $I_{E}$ is the microcanonical action, which means that the
entropy $S$ {\em {must}} be given by $S=-I_{E}$ \cite{york}.
Therefore our action (\ref{eq:IE}) requires modification in the electrically charged
case.
This has been studied in detail by Mann and Ross \cite{mann}, and here we briefly
summarize their results.
We start by considering, instead of our Euclidean instanton without boundary,
a Euclidean section with boundary (for example, one could take one half of our
Euclidean instanton, taking the boundary to be along the surface $\tau =0, \pi /2$
\cite{mann}).
In order to have a well-defined variational principle with the electric charge
fixed on the boundary, Mann and Ross found it necessary to add a boundary term
to the action (\ref{eq:IE}), and consider instead \cite{mann}:
\begin{equation}
{\tilde {I}}_{E} = I_{E} - \frac {1}{4\pi } \int _{\partial {\cal {M}}}
d^{3}x {\sqrt {h}} F^{\mu \nu }A_{\nu } n_{\mu },
\label{eq:IEtilde}
\end{equation}
where $n_{\mu }$ is the unit normal to the boundary.
In our case, we wish to compute the action for a manifold without boundary,
which would imply that the additional term in (\ref{eq:IEtilde}) makes no
difference.
However, we may write the boundary term as a volume integral:
\begin{equation}
{\tilde {I}}_{E} = I_{E}
- \frac {1}{4\pi } \int _{{\cal {M}}} d^{4}x {\sqrt {g}} \nabla _{\mu }
\left( F^{\mu \nu }A_{\nu } \right) ,
\end{equation}
which will then contribute even when there is no boundary.
On-shell, for an electromagnetic field satisfying Maxwell's equations, we have
the identity
\begin{equation}
\nabla _{\mu } \left( F^{\mu \nu } A_{\nu } \right)
= \left( \nabla _{\mu } F^{\mu \nu } \right) A_{\nu }
+ F^{\mu \nu } \nabla _{\mu } A_{\nu }
= \frac {1}{2} F^{\mu \nu }F_{\mu \nu },
\end{equation}
and so
\begin{equation}
{\tilde {I}}_{E}^{{\rm {on-shell}}} = I_{E}
- \frac {1}{8\pi } \int _{{\cal {M}}} d^{4}x {\sqrt {g}}
F^{\mu \nu }F_{\mu \nu }.
\end{equation}
The advantage of using this on-shell action is that it is manifestly gauge-invariant.
Finally, computing ${\tilde {I}}_{E}^{{\rm {on-shell}}}$, we find
\begin{equation}
{\tilde {I}}_{E}^{\rm {on-shell}} = - \pi  K^{2} l^{2} \left( 1 - \frac {2M}{l} \right),
\end{equation}
which is indeed minus the total entropy.
We note that one could instead follow precisely the calculation of \cite{mann},
and find (\ref{eq:IEtilde}) for one-half the Euclidean instanton, using
the boundary term in (\ref{eq:IEtilde}), and then multiply the final answer by two.
Note that for the boundary considered in \cite{mann}, the extrinsic curvature $K$
is zero, so that the generalized Gibbons-Hawking term in (\ref{eq:IE}) does not
contribute.

An alternative approach to the computation of the entropy for the lukewarm black holes
has been followed in \cite{cai}.
There, the authors used a result of Kallosh et al \cite{kallosh}, that the entropy
of a black hole in a general theory with arbitrary, but minimally coupled, matter
is given by
\begin{equation}
S = \frac {1}{8\pi } \int _{r_{+}}^{r_{++}}
d^{3} x {\sqrt {h}}  \left[ K -K_{0} \right] ,
\label{eq:kallosh}
\end{equation}
where $K_{0}$ is the extrinsic curvature for a reference background (in our case
de Sitter space).
The expression on the right-hand-side in (\ref{eq:kallosh}) needs some explanation.
One explicity evaluates integrals over the two 3-surfaces corresponding to the event and
cosmological
horizons, even though the Euclidean manifold has no boundaries at these points.
In \cite{cai} it is shown that this procedure yields the correct total entropy for
the lukewarm RNdS black holes without a scalar field.
With a conformally coupled scalar field, we can see from the action (\ref{eq:IE}) that
(\ref{eq:kallosh}) will become instead
\begin{equation}
S = \frac {1}{8\pi } \int _{r_{+}}^{r_{++}}
d^{3} x {\sqrt {h}} \left[ 1 - \frac {\phi ^{2}}{6} \right]
\left[ K -K_{0} \right] ,
\end{equation}
and, since the scalar field $\phi $ depends only on the radial co-ordinate $r$,
it is constant over the horizon 3-surfaces and we therefore obtain the total entropy
(\ref{eq:Stotal}).

\section{Local Horizon Energy}
\label{sec:firstlaw}

In this paper we have been considering only the region of space-time between the
event and cosmological horizons and therefore we do not consider total mass
and charge as quantities defined at infinity.
Instead, we shall use the isolated horizons formalism \cite{ashtekar} to define
local horizon energies for the event and cosmological horizons.
In this section we consider only the electrically charged black holes,
as magnetic charge does not appear in the first law for isolated horizons \cite{ashtekar1}.
Here we take a fairly simplistic view of the horizon energy, more detailed discussion
of the subtleties involved in defining this concept for black holes in de Sitter space are
discussed in \cite{corichi}.

In \cite{ashtekar}, the isolated horizons formalism is extended to include
a scalar field non-minimally coupled to the geometry.
From this, the horizon energy $E_{\Delta }$ of an isolated horizon is defined.
It satisfies the first law:
\begin{equation}
\delta E_{\Delta } = T_{\Delta } \delta S_{\Delta }
+ \Phi _{\Delta } \delta Q_{\Delta }.
\label{eq:first}
\end{equation}
Here $T_{\Delta }$ is the temperature of the horizon (\ref{eq:temp}),
$S_{\Delta }$ is the entropy,
given by the formula (\ref{eq:entropy}), $\Phi _{\Delta }$ is the electric potential
on the horizon and $Q_{\Delta }$ the horizon charge.
In our case, the horizon charge is simply $Q_{\Delta }=Q$ for both the event and cosmological
horizons, and we have
\begin{equation}
\Phi _{\Delta } = \frac {Q}{r_{\Delta }}.
\end{equation}
However, in order to implement (\ref{eq:first}), we first note that we are dealing with
a restricted phase space, and cannot regard $Q$ as an independent variable, since it
is constrained to satisfy (\ref{eq:qm})
by the coupling constants in the theory.
Therefore, we must set $\delta Q_{\Delta } = K \delta M$, where $K$ is fixed.

\begin{figure}
\begin{center}
\includegraphics[angle=270,width=10cm]{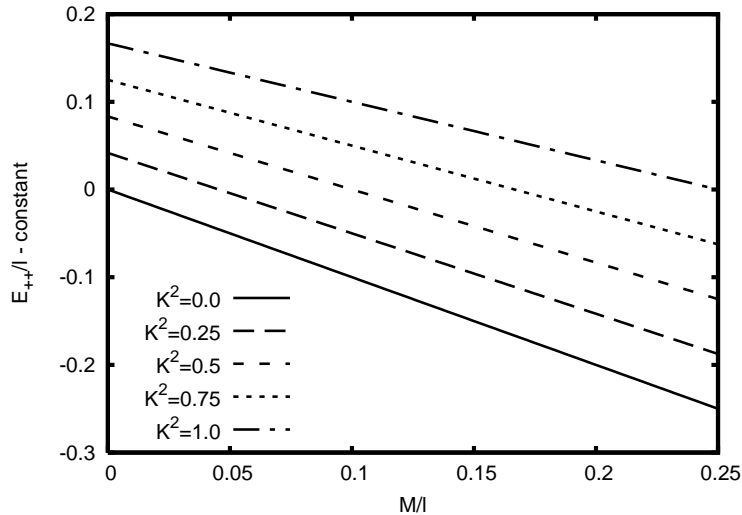}
\end{center}
\caption{Graph of the cosmological horizon energy $E_{++}$ as a function of the
parameter $M$
for various values of the charge/mass ratio.}
\label{fig:E++}
\end{figure}

Considering firstly the black hole event horizon, taking the variation of the
formula for the entropy (\ref{eq:Sr+}), and substituting in (\ref{eq:first}), we find
\begin{equation}
\delta E_{+} = \delta M ,
\end{equation}
from which we deduce that $E_{+}=M$ up to a constant.
This result is not unexpected, since, from the form of the metric (\ref{eq:metric}) we
would regard $M$ as being the ``mass'' of the black hole.
However, performing the same calculation for the cosmological horizon yields instead
\begin{equation}
\delta E_{++} =
\left[ -1 + K^{2} \left( 1 - {\sqrt {1-\frac {4M}{l}}} \right) \right]
\delta M ,
\end{equation}
so that
\begin{equation}
E_{++} = \left( K^{2} -1 \right) M + \frac {K^{2}l}{6}
\left( 1 - \frac {4M}{l} \right) ^{\frac {3}{2}}
+ {\mbox {constant}} .
\end{equation}
The cosmological horizon energy therefore includes contributions from the matter fields
(scalar and electromagnetic fields) between the horizons.
There are two limits of particular interest.
Firstly, for RNdS black holes without a scalar field, we have $K=1$ and
\begin{equation}
E_{++} = \frac {l}{6}
\left( 1 - \frac {4M}{l} \right) ^{\frac {3}{2}}
+ {\mbox {constant}} ;
\label{eq:RNE++}
\end{equation}
and secondly there are the neutral black holes for which $K=0$ and
\begin{equation}
E_{++} = -M + {\mbox {constant}} .
\end{equation}
In both these cases, we observe that the cosmological horizon energy decreases as the
parameter $M$ increases.
In fact, it is straightforward to see that $E_{++}$ is decreasing as $M$ increases
for all values of the charge/mass ratio $K$, as shown in Figure \ref{fig:E++}.
This is in accord with pure de Sitter space having larger energy than any asymptotically
de Sitter space-time.

For the RNdS black holes without a scalar field, the region outside the cosmological
horizon has been studied in \cite{radu}, with the gravitational mass and total charge
computed using a background counterterm technique.
As may be expected from the form of our particular metric (\ref{eq:metric}),
the total charge is found to be $M$.
The gravitational mass is found to be $-M$, so the mass of the space-time is less
than that of de Sitter space.
Compared with our expression for the energy of the cosmological horizon (\ref{eq:RNE++}),
we see that the electromagnetic field outside the cosmological horizon makes
a significant contribution to the total gravitational mass.
One would like to perform a similar calculation including the scalar field, but
this will necessitate the formulation of suitable counterterms for the scalar field
contribution to the action.
This issue will be addressed elsewhere \cite{E+R}.

\section{Conclusions}
\label{sec:conc}

In this paper we have studied the classical thermodynamics of the black hole solutions
of Martinez, Troncoso and Zanelli \cite{mtz}, in which there is a de Sitter black hole
with a conformally coupled scalar field.
Unlike the corresponding flat space black hole \cite{BBMB}, in this case we have a
finite, non-zero, temperature and entropy.
For most of the black holes, we find that the entropy of the event horizon is negative,
but the entropy of the cosmological horizon and the total entropy are always
positive.
The total entropy is found to be equal to minus the Euclidean action, as anticipated,
but there is a subtlety in the computation in the electrically charged case.
Following Mann and Ross \cite{mann}, and including an additional boundary
term to fix the charge on the boundary, gives the correct answer.
Finally, we computed the local horizon energies of the event and cosmological horizons,
using the first law as derived from the isolated horizons formalism \cite{ashtekar}.
Once again, the non-minimal coupling of the scalar field has a significant effect.
While the energy of the event horizon is the mass parameter $M$ appearing in the metric,
the energy of the cosmological horizon is more complicated, as it has contributions
from the matter fields between the horizons.
It will be of interest to compare these local energies with
the gravitational mass at infinity, computed using counterterm
subtraction techniques, and we hope to return to this question in the near future.
It would also be instructive to compare the thermodynamics of these de Sitter black holes
with the corresponding behaviour of the anti-de Sitter black holes with a conformally coupled
scalar field \cite{ew}.

\begin{acknowledgments}
We would like to thank Alejandro Corichi, Eugen Radu and Phil Young for helpful discussions.
EW would like to thank the Department of Mathematical Physics, University College Dublin,
for hospitality while this work was completed.
This work was supported by UK PPARC, grant reference number PPA/G/S/2003/00082,
the Royal Society and the London Mathematical Society.
\end{acknowledgments}

\end{document}